\documentclass[aps,showpacs,pra,preprint,amsmath,amssymb,superscriptaddress]{revtex4-1}
\usepackage{bm}
\usepackage{amssymb}
\usepackage{colordvi}
\usepackage{graphicx}
\usepackage{color}
\usepackage{hyperref}

\newcommand{\be}{\begin{equation}}
\newcommand{\ee}{\end{equation}}
\newcommand{\bea}{\begin{eqnarray}}
\newcommand{\eea}{\end{eqnarray}}
\newcommand{\ep}{\epsilon}
\newcommand{\vep}{\varepsilon}
\newcommand{\ave}[1]{\langle #1\rangle}
\newcommand{\ome}{\omega}
\newcommand{\Ome}{\Omega}

\def\ket#1{\vert #1 \rangle}

\begin{document}

\title{Quantum many-body simulation using monolayer exciton-polaritons
  in coupled-cavities}

\author{Hai-Xiao Wang}
\affiliation{College of Physics, Optoelectronics and Energy, \&
  Collaborative Innovation Center of Suzhou Nano Science and
  Technology, Soochow University, 1 Shizi Street, Suzhou 215006,
  China}
\author{Alan Zhan}
\affiliation{Department of Physics, University of Washington, Seattle, Washington 98195, USA}
\author{Ya-Dong Xu}
\affiliation{College of Physics, Optoelectronics and Energy, \&
  Collaborative Innovation Center of Suzhou Nano Science and
  Technology, Soochow University, 1 Shizi Street, Suzhou 215006,
  China}
\author{Huan-Yang Chen}
\affiliation{College of Physics, Optoelectronics and Energy, \&
  Collaborative Innovation Center of Suzhou Nano Science and
  Technology, Soochow University, 1 Shizi Street, Suzhou 215006,
  China}
\author{Wen-Long You}\email{wlyou@suda.edu.cn}
\affiliation{College of Physics, Optoelectronics and Energy, \&
  Collaborative Innovation Center of Suzhou Nano Science and
  Technology, Soochow University, 1 Shizi Street, Suzhou 215006,
  China}
\author{Arka Majumdar}\email{arka@uw.edu}
\affiliation{Department of Physics, University of Washington, Seattle, Washington 98195, USA}
\affiliation{Department of Electrical Engineering, University of
  Washington, Seattle, Washington 98195, USA}
\author{Jian-Hua Jiang}\email{jianhuajiang@suda.edu.cn}
\affiliation{College of Physics, Optoelectronics and Energy, \&
  Collaborative Innovation Center of Suzhou Nano Science and
  Technology, Soochow University, 1 Shizi Street, Suzhou 215006,
  China}

\date{\today}

\begin{abstract}
Quantum simulation is a promising approach to understand complex
strongly correlated many-body systems using relatively simple and
tractable systems. Photon-based quantum simulators have  great
advantages due to the possibility of direct measurements of
multi-particle correlations and ease of simulating non-equilibrium
physics. However, interparticle interaction in existing photonic
systems is often too weak limiting the potential of 
quantum simulation. Here we propose an approach to enhance the interparticle
interaction using exciton-polaritons in MoS$_2$ monolayer quantum-dots
embedded in 2D photonic crystal microcavities. Realistic calculation
yields optimal repulsive interaction in the range of $1$-$10$~meV ---
more than an order of magnitude greater than the state-of-art
value. Such strong repulsive interaction is found to emerge neither in
the photon-blockade regime for small quantum dot nor in the
polariton-blockade regime for large quantum dot, but in the crossover
between the two regimes with a moderate quantum-dot radius around
20~nm. The optimal repulsive interaction is found to be largest in
MoS$_2$ among commonly used optoelectronic materials. Quantum
simulation of strongly correlated many-body systems in a finite chain
of coupled cavities and its experimental signature are studied via
exact diagonalization of the many-body Hamiltonian. A method to
simulate 1D superlattices for interacting exciton-polariton gases in
serially coupled cavities is also proposed. Realistic considerations on
experimental realizations reveal advantages of transition
metal dichalcogenide monolayer quantum-dots over conventional
semiconductor quantum-emitters.
\end{abstract}


\maketitle

\section{Introduction}
Solving strongly correlated quantum many-body
systems exactly is a formidable task. One promising approach is to
mimic such complicated systems using another simpler and easily
controllable quantum system, as envisioned by
Feynman\cite{feynman}. To that end, the first 
demonstration of quantum phase transition with ultracold atoms in an
optical lattice sparked a large body of research on quantum
simulation in ultracold atomic systems\cite{nat,natphys}. Interacting
photons also provide a unique and
distinctive platform to study strongly correlated quantum many-body
systems\cite{lukin,CCA_1,CCA_2,CCA_3,CCA_4}. The main idea behind this 
approach is to create an interacting ``quantum fluid of
light''\cite{ciutirmp} via a coupled network of nonlinear photonic
cavities\cite{lukin,CCA_1,CCA_2,CCA_3,CCA_4,jelena,ciutirmp,polariton}. 
Advantages of photonic quantum simulators include
much higher energy scale and faster operations, available
non-destructive techniques for direct measurements of quasiparticle
properties via spatial- and/or
time-resolved multi-photon correlation functions, and abundant optical methods for coherent
control\cite{ciutirmp,3rdg}. Such multi-particle correlation measurement is extremely difficult in
both cold-atomic gases and strongly correlated electronic materials.
These advantages yield great promises for photonic quantum many-body
simulation as a way to understand the role of many-body quantum
entanglement in Mott insulators which remains an outstanding challenge
to fundamental physics\cite{wen}.

Polariton, a quantum superposition of a photon and an exciton, emerges in
hybrid strongly coupled systems of photonic microcavity and
semiconductor excitons \cite{polariton,prx}. The composite nature of 
polaritons leads to various unusual properties, such as high-temperature
Bose-Einstein condensation (BEC)\cite{cdte,snoke,prx,sr}, and enhanced
optical nonlinearity for applications in all-optical
diodes\cite{diode} and transistors\cite{transistor}. However,
optical nonlinearity in those systems generally requires high
polariton densities. Achieving optical nonlinearity at the single
photon level requires significant reduction of the cavity
mode-area/volume and optimization of the optoelectronic material
(typically forming a semiconductor quantum dot)\cite{verger}. Note that, such
single photon nonlinearity is necessary to realize the
aforementioned photonic quantum simulators. Photon blockade, the
effect where a single photon repels other photons, has been observed
using a very small quantum dot (QD) coupled to a
cavity\cite{pblock0,pblock1,pblock2,ele2}. In those systems,  
Pauli blockade forbids double-occupancy of excitons, hence
the interaction between polaritons is simply given by the energy
difference between free polaritons and the Pauli blockade polaritons,
i.e., $U_{pl}=(2-\sqrt{2})\hbar\Ome$, where $\hbar\Ome$ denotes the
exciton-photon interaction
strength\cite{pblock0,pblock1,pblock2}. However, the 
area of QD $\sim (10~nm)^2$, is much smaller than the modal area of
the optical cavity, leading to much reduced light-matter
interaction and polariton-polariton repulsion. The state-of-the-art
value of polariton repulsive interaction in the photon-blockade regime
is less than 0.1~meV\cite{ele2}. Thus an important challenge for
polariton quantum many-body simulation is to realize much stronger
repulsive interaction. 

In this work, we propose an optoelectronic architecture to realize
polariton repulsion much larger than the state-of-the-art value,
1$\sim$10~meV, using exciton-polaritons based on monolayer MoS$_2$ QDs
embedded in slab photonic crystal cavities. The strength of the
repulsive interaction varies with the quantum dot radius due to
the competition between the exciton-photon interaction and the
exciton-exciton repulsion. It is found that the strongest repulsive
interaction emerges neither in the photon-blockade regime for small
QDs nor in the polariton-blockade regime for large QDs, but in the
crossover between the two regimes. An optimal quantum dot radius is
found as $\sim 20$~nm. Similar trends are found for other common
materials such as GaAs, InAs, CdTe,
and GaN. Nevertheless, MoS$_2$ provides the largest 
nonlinearity, thanks to strong light-matter and
exciton-exciton interactions. We further investigate possible
experimental consequences of quantum simulation in a chain of coupled
cavities using exact diagonalization of the many-body
Hamiltonian. In addition, a method for simulation of superlattices in
coupled cavities is proposed and the regimes for Mott transition is
estimated using single- and two-particle analysis. Realistic
considerations for fabrication and measurements reveal
advantages of transition metal dichalcogenide (TMD) monolayer
semiconductors over conventional optoelectronic materials.

\begin{figure}[]
\begin{center}
\includegraphics[height=4.7cm]{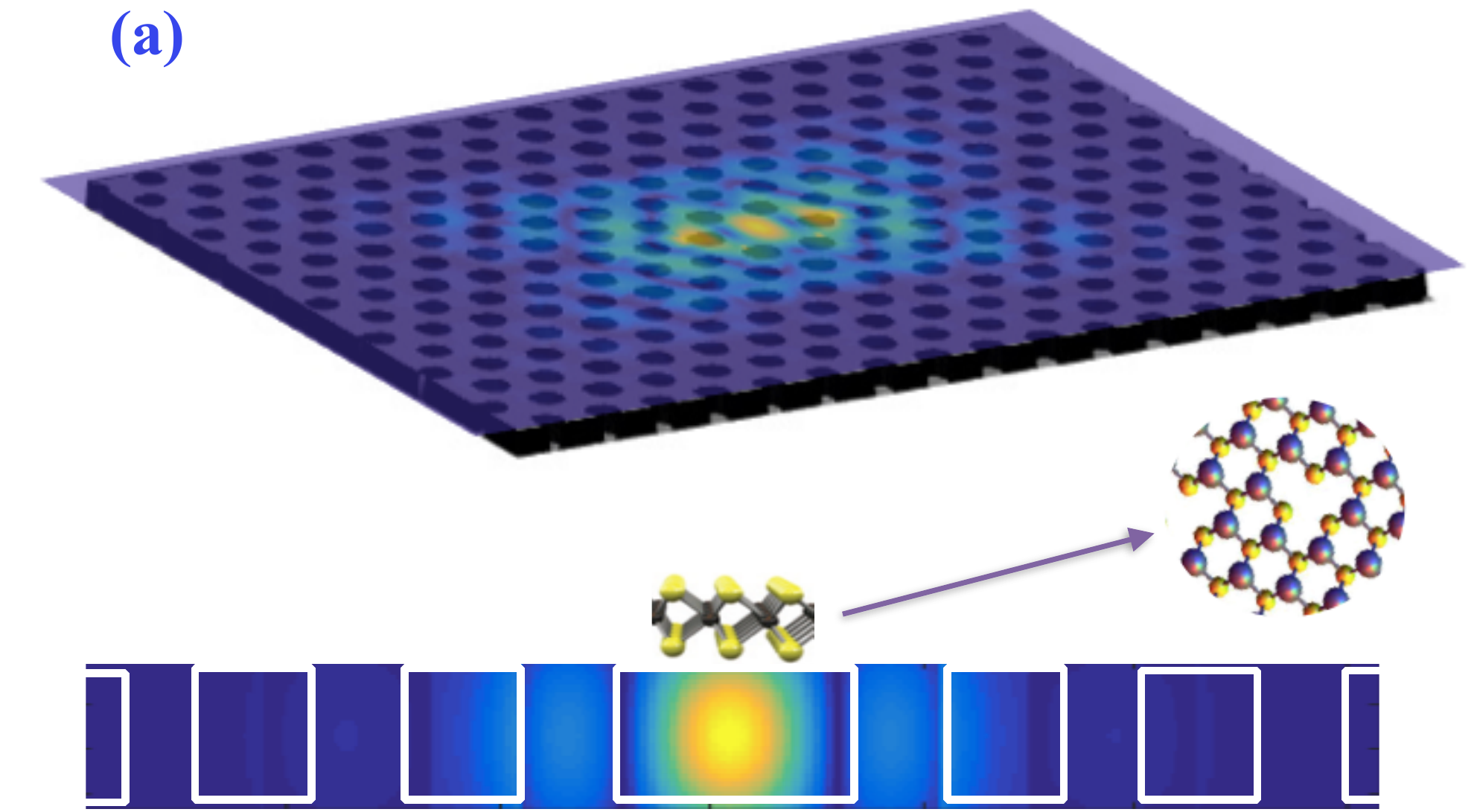}\includegraphics[height=5.cm]{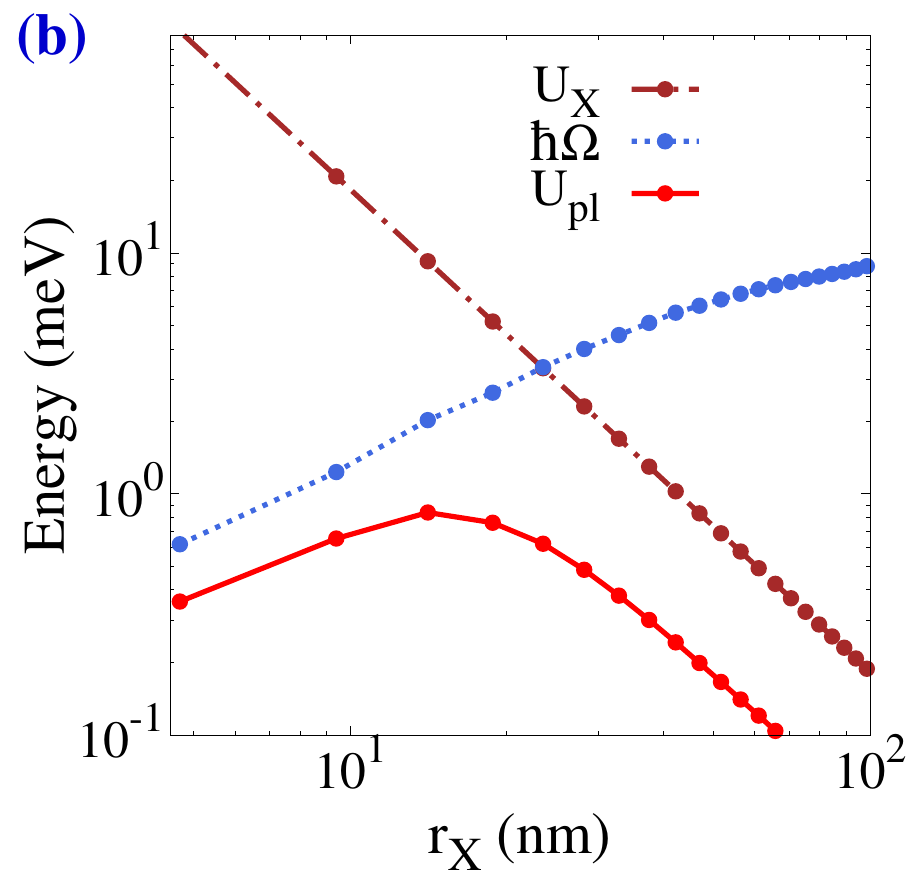}
\caption{(a) Cavity-QD hybrid system for strongly
  interacting polaritons. Upper panel: The H1 cavity will be realized
  using a thin 2D photonic crystal slab. A MoS$_2$ QD is placed
  at the center of the cavity. Lower panel:
  Inplane electric field distribution in the $x$-$z$ plane (field
  outside the membrane is not plotted). The position of the MoS$_2$ QD
  is illustrated using a schematic of atomic structure 
  of MoS$_2$. (b) Exciton-exciton interaction $U_X$, exciton-photon
  interaction $\hbar\Ome$, and polariton-polariton interaction
  $U_{pl}$ as functions of the QD radius $r_X$ for MoS$_2$ with zero
  detuning. }
\label{fig1}
\end{center}
\end{figure}

\section{Material and Photonic Architecture}
The proposed architecture is illustrated in Fig.~1(a). The cavity is
formed by a point defect in a 2D hexagonal photonic crystal slab, also
known as the H1 cavity \cite{ourcavity} [see Supplementary
Materials\cite{SM}]. The MoS$_2$ QD can be fabricated via patterning a
MoS$_2$ monolayer and placed at the 
center of the cavity. A network of the cavity-QD hybrid structure
forms an interacting polariton lattice system, which 
is described by the following Hamiltonian\cite{ciutirmp}
\begin{align}
& {\cal H} = \sum_{i} \big [\hbar \ome_{c} c_{i}^\dagger c_{i} + \hbar \ome_X b_{i}^\dagger
b_{i} + \hbar \Ome (c_{i} b_{i}^\dagger + b_{i} c_{i}^\dagger ) 
+  \frac{1}{2} U_X {\cal N}_i ({\cal N}_i-1) 
\big] - t \sum_{\langle i,j\rangle} c_{i}^\dagger c_{j}
. \label{Hamil} 
\end{align}
Here $\ome_{c}$ is the frequency of the cavity mode, $c_{i}^\dagger$
($b_{i}^\dagger$) creates a photon (exciton) in the $i^{th}$ cavity
(QD), ${\cal N}_i= b_{i}^\dagger b_{i}$ stands for the exciton number
operator, $U_X$ denotes the exciton-exciton repulsion, $\hbar\Omega$
represents the exciton-photon interaction, and $\hbar\ome_X=1.87$~eV
is the exciton energy in MoS$_2$ QDs. We assume that $\ome_c$ is
identical for each cavity and $\ome_X$ is the same for each MoS$_2$
QD. The effects of fluctuation and disorder will be considered
later. From Refs.~\cite{prx,sr}, the exciton-photon coupling is 
\begin{align}
\hbar\Ome &= \frac{d_{cv}|\phi(0)|\sqrt{\hbar\ome_{c}
    }}{\sqrt{2\ep_0 L_c }} , \label{lc-ome}
\end{align}
where $d_{cv}=4.0\times 10^{-29}$~C$\cdot$m is the interband dipole matrix
element\cite{sr,nc-mse2} and $|\phi(0)|=\sqrt{2/(\pi a_B^2)}$ is 
the exciton wave amplitude at zero electron-hole distance ($a_B=1$~nm
is the exciton Bohr radius in MoS$_2$\cite{diana}). The exciton-photon
coupling depends on the following quantity of the dimension of length,
\begin{align}
& L_c \equiv \frac{ \int_{c} d{\vec r} \ep({\vec r}) |{\vec E}({\vec r})|^2}{\int_c dxdy |{\vec E}(x,y,z_0)|^2\Theta(x,y,z_0)} ,\label{lc-area}
\end{align}
where $\ep({\vec r})$ is the position-dependent (relative) dielectric
constant, ${\vec E}(\vec{r})$ is the electric field of the cavity
mode, and $z_0$ is the $z$ coordinate of the MoS$_2$ monolayer. The
$\Theta(x,y,z_0)$ function, which takes into account the finite
overlap between the QD and the cavity optical field, is unity in the QD
region and zero outside\cite{prx}. The integrals are carried out within
each cavity. The exciton-exciton interaction strength is given by\cite{tassone}
$U_X = \frac{6 E_b a_B^2}{S_X}$,
where $E_b=0.96$~eV is the exciton binding energy and $S_X=\pi r_X^2$ is the
area of the circular MoS$_2$ QD with radius $r_X$. 
The last term in Eq.~(\ref{Hamil})
describes photon hopping between nearest-neighbor cavities, where $t$ is the hopping energy.
Note that in the above formalism, the exciton-polariton is
approximately treated as uniformly distributed in the QDs [resulting in
$\Theta(x)$ in Eq.~(\ref{lc-area}) and the $S_x$ factor in
$U_X$]. More rigorous treatment with non-uniform distribution is
equivalent to a correction of the effective area of the polariton,
which affects the results marginally [see supplementary materials].

The designed H1 cavity has a slab thickness of 110~nm and a lattice
periodicity of  $a=190$~nm to ensure that the fundamental TE mode is
resonant with the MoS$_2$ exciton ($\lambda_X=660$~nm; $\lambda_X$ is the
photon wavelength in vacuum for frequency $\ome_X$). Gallium
phosphide is chosen as the material for the slab photonic crystal
cavity, due to its high refractive index $(n=3.2)$ and transparency in
that wavelength range. The choice of H1 cavity is primarily motivated
by its small mode-volume $(\sim 0.45(\lambda_X/n)^3)$ and mode area $(\sim
(\lambda_X/n)^2)$.

\section{Effective Hamiltonian and Polariton-polariton interaction}
In the uncoupled limit, photon is itinerant and exciton is
localized. All interesting physics comes in when the light-matter
interaction is turned on. In the regime when the light-matter
interaction $\hbar\Ome$ is much greater than the photon hopping
$t$\cite{CCA_3}, the many-body quantum dynamics close to the ground
state is constrained to the lower-polariton Hilbert space and one can
truncate the full Hamiltonian (\ref{Hamil}) into the following
effective Hamiltonian \cite{ciutirmp}
\be
{\cal H}_{pl} = - t_{pl} \sum_{<i,j>}
a_{i}^\dagger a_{j} + \frac{1}{2} \sum_{i} U_{pl} n_i (n_i -1) . \label{pol-in}
\ee
Here $t_{pl}= t p_c$ and $n_{i}=a_{i}^\dagger a_{i}$ with
$a_{i}^\dagger$ being polariton creation operator. $p_c \equiv
\cos^2[\frac{1}{2}{\rm arccot}(\frac{\Delta}{2\hbar\Ome})]$ is the
photonic fraction of the lower polariton\cite{ciutirmp}, where
$\Delta\equiv \hbar(\ome_X-\ome_c)$ is the exciton-photon detuning. The
polariton-polariton interaction $U_{pl}$ is determined by the
difference between the ground state energy of an isolated cavity with
two quanta with and without the exciton-exciton repulsion,
respectively\cite{CCA_3}, 
\be
U_{pl}\equiv E_{GS}(2q)-E^{(0)}_{GS}(2q) .  \label{upl}
\ee
The ground state energy of the polaritonic system is calculated based
on the following: The Hamiltonian of an isolated cavity with two
energy quanta can be written in the basis of
$(\ket{2,0},\ket{1,1},\ket{0,2})^T$ (here $\ket{n_p,n_x}$
with $n_p+n_x=2$ are the Fock states with $n_p$ photons and $n_x$
excitons) as
\be
{\cal H}_{2q} = \left(\begin{array}{cccc} 2\hbar\ome_c & \sqrt{2}\hbar\Ome & 0  \\
    \sqrt{2}\hbar\Ome & \hbar(\ome_c + \ome_X) & \sqrt{2}\hbar\Ome \\
    0 & \sqrt{2}\hbar\Ome & 2\hbar\ome_X + U_X \\
\end{array}\right) .  \label{2qe}
\ee
The ground state of the above Hamiltonian consists of two interacting
polaritons, of which the total energy is $E_{GS}(2q)$. When the
interaction between exciton is turned off, $U_X=0$, the ground state
of the Hamiltonian gives two noninteracting polaritons, with total
energy $E^{(0)}_{GS}(2q)$. The difference between the two energies of the
ground states is the interaction energy between two polaritons within
a cavity.

In the literature, there are two distinct regimes in which photon
antibunching were observed and studied: (i) the photon blockade
regime\cite{pblock0,pblock1,pblock2,ele2} where the QD size is 
small and thus $U_X\gg \hbar\Ome$, (ii) the polariton blockade 
regime\cite{verger} where the QD size is large and then $U_X\ll
\hbar\Ome$. Photon blockade was observed experimentally in cavity-QD
hybrid systems using small InAs QDs where the value of exciton-photon
coupling strength is small, $\hbar\Ome\le 0.16$~meV\cite{ele2}. In
these systems the polariton interaction $U_{pl}$ is weak, $U_{pl}\le
0.1$~meV\cite{ele2}.

One of the main conclusions in this paper is that the maximum
polariton-polariton interaction is not reached in the photon blockade
regime where the exciton-exciton 
repulsion is very strong, nor in the polariton blockade regime where
the light-matter interaction is very strong. As illustrated in
Fig.~1(b), the polariton-polariton interaction $U_{pl}$ ramps up when
the QD radius $r_X$ is small (the photon blockade regime). After
reaching to a maximum value around $r_X=20$~nm, $U_{pl}$ decays with
the QD radius in the polariton blockade regime. The maximum value of
$U_{pl}$ lies in the crossover between the two regimes. To the best of
our knowledge such non-monotonic behavior (also holds for other
materials, see Supplementary Materials\cite{SM}) is never reported
before. This finding indicates that there is an optimal QD radius 
for strong polariton-polariton interaction in each optoelectronic
material.

\begin{figure}[]
\begin{center}
\includegraphics[height=5.cm]{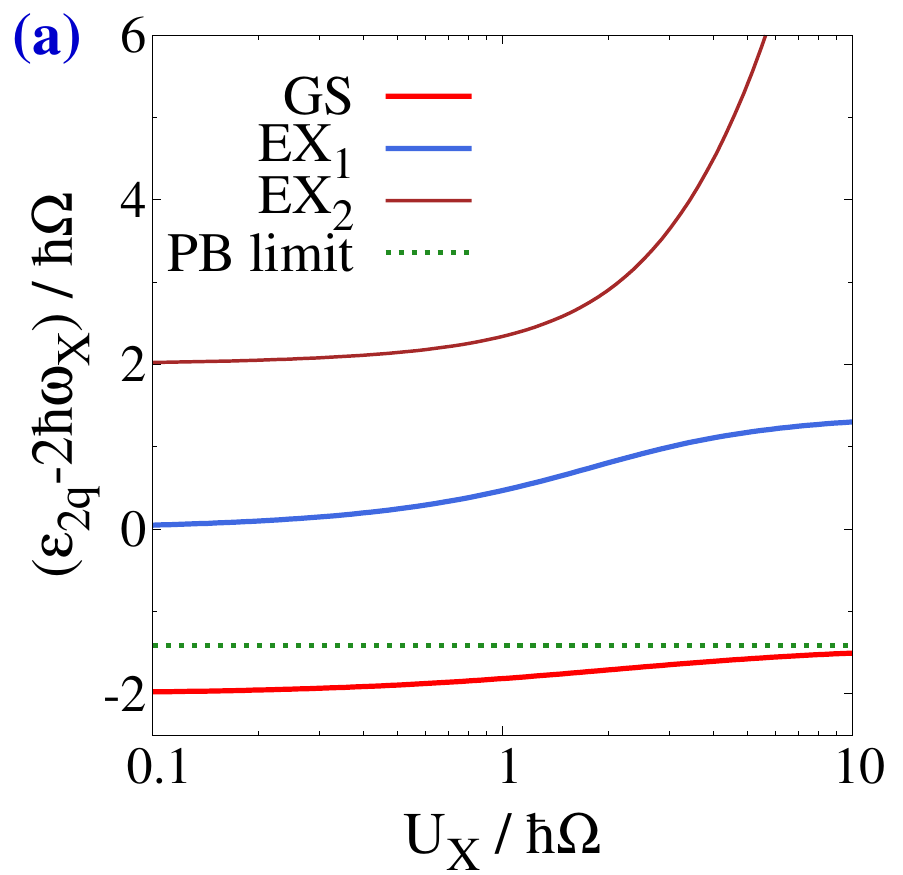}\includegraphics[height=5.cm]{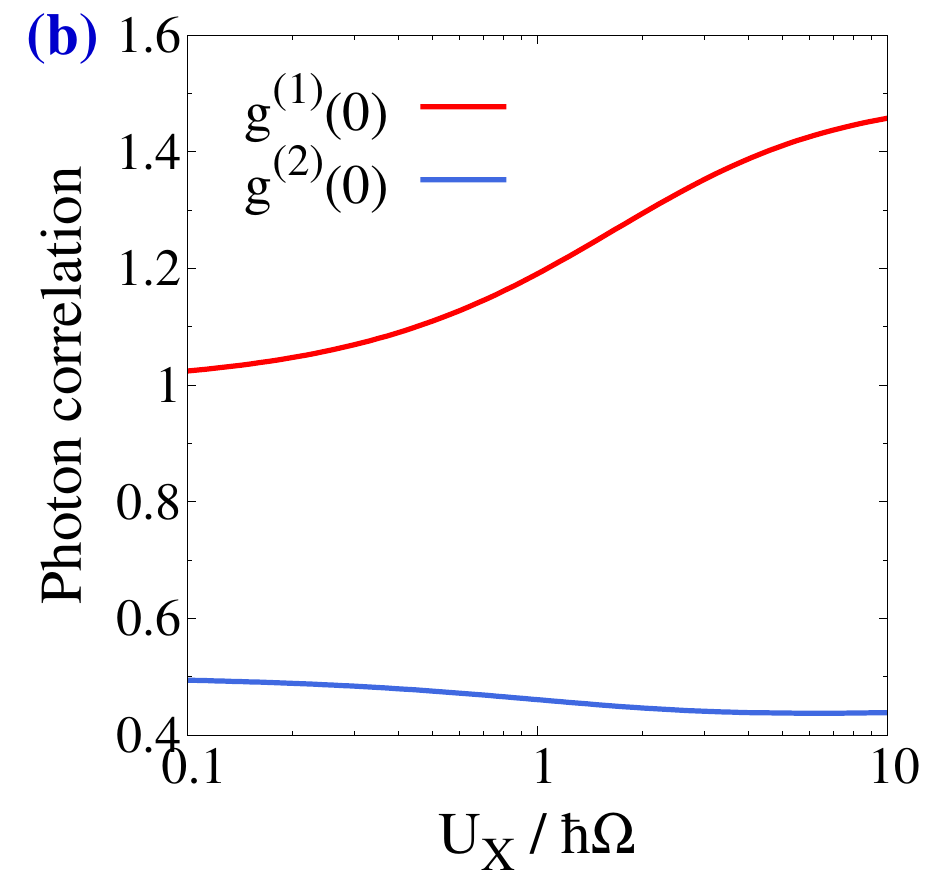}
\includegraphics[height=5.cm]{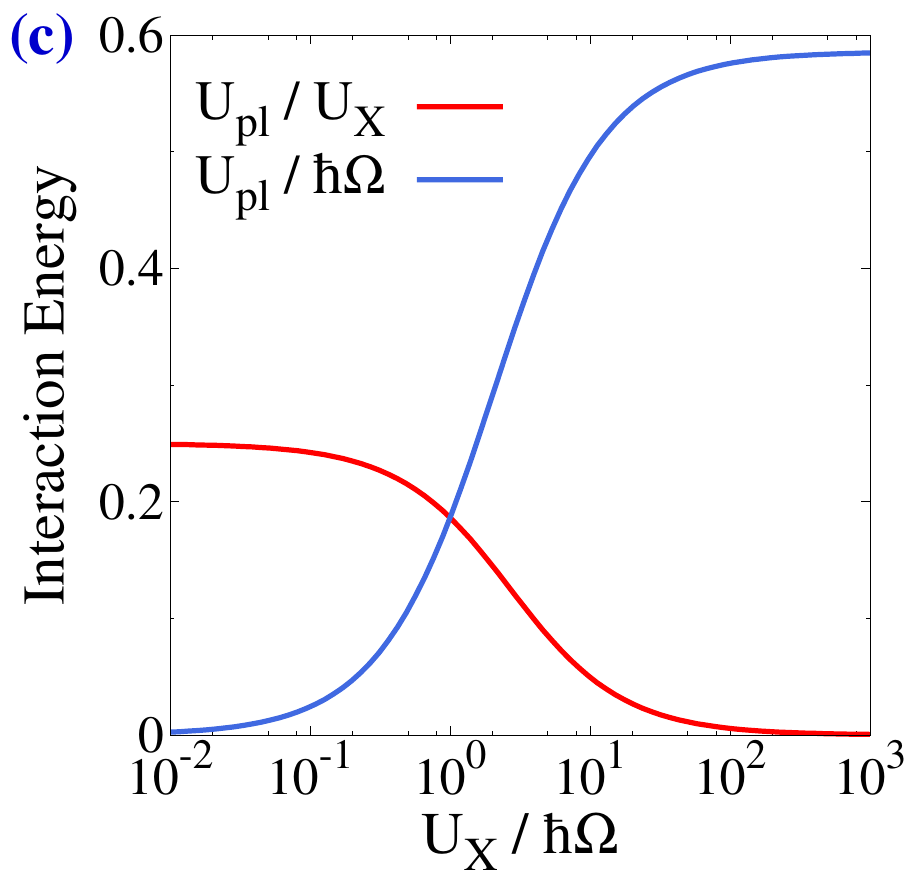}\includegraphics[height=5.cm]{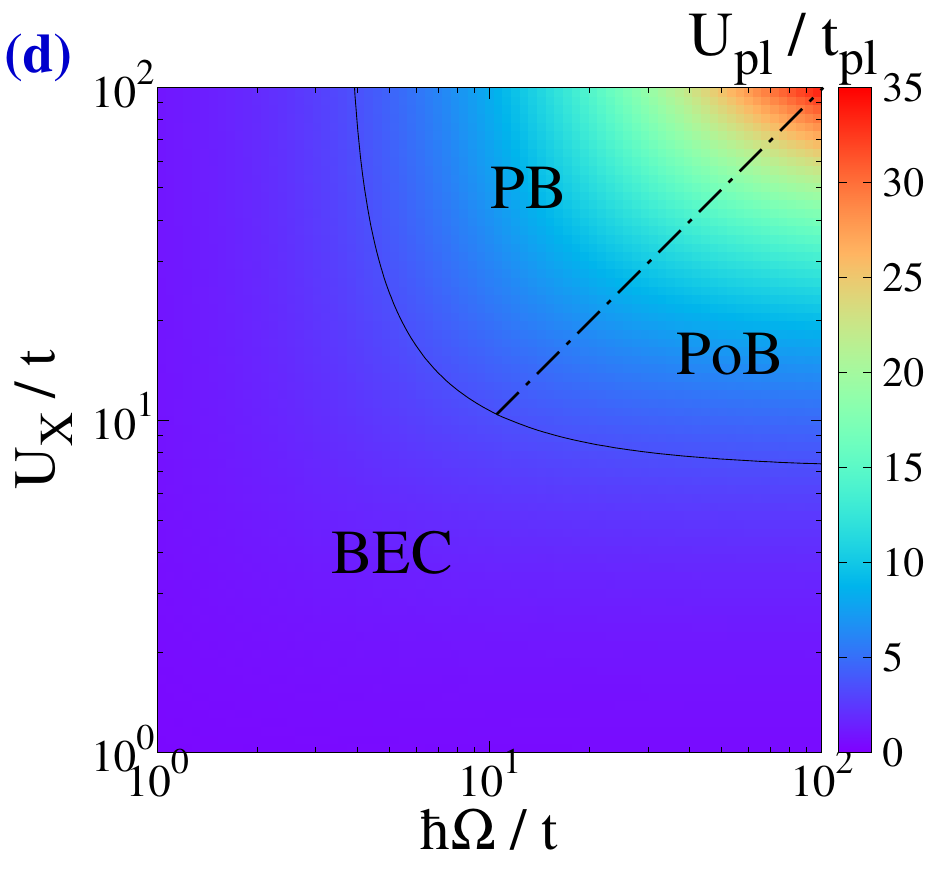}
\caption{ Crossover from polariton blockade to photon blockade. (a)
  Energy levels $\vep_{2q}$ of the ground state (GS), the first and
  second excited states (EX$_1$ and EX$_2$) as functions of the
  exciton-exciton interaction $U_X$ for a single cavity with two
  energy quanta. We scale the interaction energy $U_X$ with the
  exciton-photon coupling $\hbar\Ome$. The dotted line denote the
  energy $-\sqrt{2}\hbar\Ome$, i.e., the GS energy in the photon
  blockade (PB) limit. (b) Photon correlation function $g^{(1)}(0)$
  and $g^{(2)}(0)$ as functions of $U_X/(\hbar\Ome)$. (c) Polariton
  interaction energy $U_{pl}$ for various $U_X/(\hbar\Ome)$. (d)
  $\alpha$ and phase diagram of 1D interacting polaritons with zero
  detuning. The Mott insulator phase consists two regions: the photon
  blockade (PB) region and the polariton blockade (PoB) region. The phase
  boundary between BEC and Mott insulator is labeled by the solid curve, while
  the crossover between PB and PoB regions is labeled by the chained curve.}
\label{fig2}
\end{center}
\end{figure}

To understand the underlying physics, we calculate the spectrum and
photon correlation of the isolated cavity with two energy quanta. The
energy levels of the Hamiltonian Eq.(\ref{2qe}), denoted as
$\vep_{2q}$, as functions of the repulsive interaction $U_X$ are given
in Fig.~\ref{fig2}(a). The photon blockade limit (i.e., when exciton
repulsion $U_X$ is much larger than the exciton-photon coupling
$\hbar\Ome$) is represented by the dotted line. We find that the
ground state energy $E_{GS}(2q)$ indeed increases with exciton-exciton
repulsion $U_X$ [see Fig.~2(a)] and the photon antibunching is
stronger in the strong exciton repulsion regime [see Fig.~2(b)]. In
this regime the first-order correlation function becomes greater than
unity, as the manifestation of the projection out of the double
exciton state [i.e., the 2$^{nd}$ excited state in Fig.~2(a)] due to
strong exciton repulsion.

However, as shown in Fig.~1(b), strong exciton-exciton repulsion
requires very small QD radius. Unfortunately for such small QD the
light-matter interaction $\hbar\Ome$ is very small (due to much
reduced overlap between the exciton and the photon field) and the
polariton-polariton repulsive interaction is determined by
$U_{pl}\simeq (2-\sqrt{2})\hbar\Ome$ [see Fig.~2(c)]. Thus the
polariton interaction $U_{pl}$ is rather weakened with decreasing QD
size in this regime, as indicated in Fig.~1(b).

\begin{figure}
\begin{center}
\includegraphics[height=5.cm]{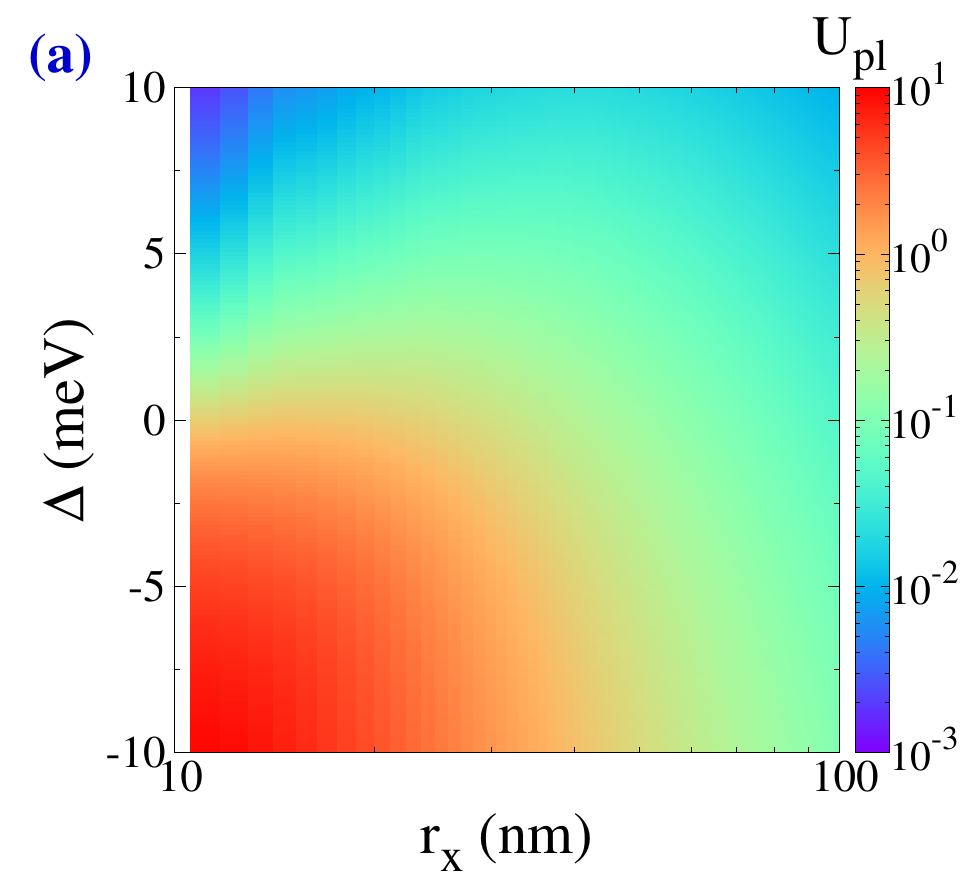}\includegraphics[height=5.cm]{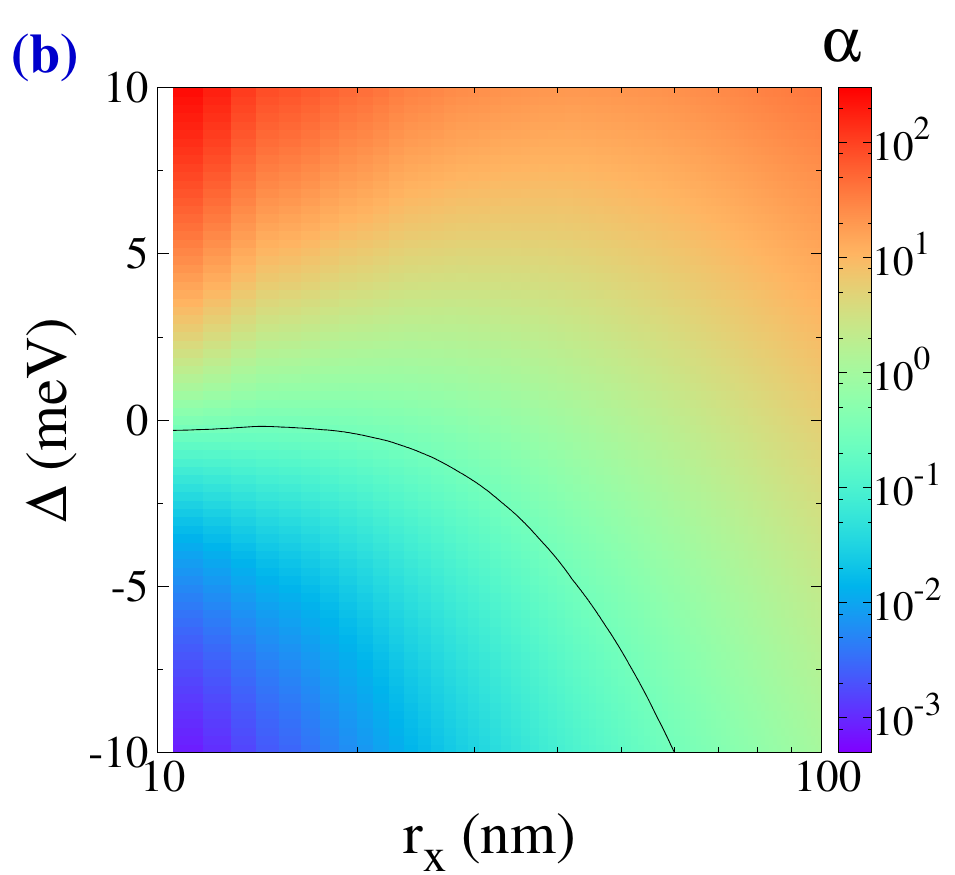}
\caption{ (a) Polariton-polariton interaction $U_{pl}$ vs. $r_X$ and
  exciton-photon detuning $\Delta$. (b) $\alpha$ and phase diagram for strongly interacting
  1D polariton system for various MoS$_2$ QD radius $r_X$ and
  exciton-photon detuning $\Delta$ with hopping energy $t=0.5$~meV.}
\end{center}
\end{figure}

In the other limit, when the light-matter interaction $\hbar\Ome$ is
much stronger than the exciton-exciton repulsion $U_X$, the polariton
interaction $U_{pl}$ is limited by the exciton-exciton repulsion
$U_X$ [see Fig.~2(c)]. In this regime, increasing the QD size leads to
greater light-matter interaction $\hbar\Ome$ but reduced exciton
repulsion $U_X$ (since $U_X=6E_ba_B^2/S_X$). Therefore, the polariton
interaction $U_{pl}$ decreases with increasing QD size. Following
these reasonings, the non-monotonic dependence 
of the polariton-polariton interaction $U_{pl}$ on the QD size shown
in Fig.~1(b) is an universal behavior for all
quantum-emitters. Indeed, we find that this behavior holds true for
other common quantum emitters, such as GaAs, InAs, CdTe, GaN and
MoSe$_2$ QDs. The optimal polariton-polariton interaction for all
these quantum-emitters obtained from the parameters for the exciton
and the light-matter interaction (i.e., $d_{cv}$, $|\phi(0)|$, $L_c$,
$E_b$, and $a_B$) are listed in Table~I, where the optimal QD
radius ranging from several nanometers to tens of nanometers [see
Supplementary Materials\cite{SM}]. We find that MoS$_2$ is one of the
best material for strong polariton-polariton interaction. The other
TMD material, MoSe$_2$, is also very appealing for quantum
simulation. Further enhancement of the polariton-polariton repulsion
can be achieved by engineering the cavity for stronger light trapping
(i.e., smaller mode area to $\lambda_X^2$ ratio; $\lambda_X$ is the
photon wavelength in vacuum for frequency $\ome_X$)

\begin{table}[htb]
  \caption{Optimal polariton-polariton interaction (in unit of meV)
    for quantum-emitters made of different materials in H1
    cavity at zero exciton-photon detuning.}
  \begin{tabular}{lllllllllllllllll}\hline \hline
    Materials & \mbox{} & MoS$_2$ & \mbox{} & MoSe$_2$ & \mbox{} & GaAs &
    \mbox{} & InAs & \mbox{} & GaN & \mbox{} & CdTe \\
    \hline
    Optimal $U_{pl}$ & \mbox{} & 0.85 & \mbox{} & 0.48 & \mbox{} &
    0.32 & \mbox{} & 0.16 & \mbox{} & 0.51 & \mbox{} & 0.33 \\
    \hline \hline
  \end{tabular}
  \label{table1}
\end{table}

We now illustrate the phase diagram of the 1D interacting polariton
system at zero exciton-photon detuning in Fig.~2(d). The polariton
Mott insulator phase exists in the region with {\em simultaneous}
strong exciton-exciton interaction and strong exciton-photon
interaction. The whole region can be separated into two regimes: the
polariton-blockade regime and the photon-blockade regime. The
crossover line (the dot-dashed line) is determined by $U_X=\hbar\Ome$.
In the other regions the polariton-polariton
interaction $U_{pl}$ is not strong enough to drive the Mott
transition, hence the system is in the BEC phase of polaritons. The
Mott-BEC phase boundary is evaluated approximately via
$\alpha=\alpha_c$ ($\alpha\equiv t_{pl}/U_{pl}$) with $\alpha_c=0.28$
for filling factor $\nu=1$ (i.e., one polariton per
cavity)\cite{PhysRevB.58.R14741}.

Using the material parameters of MoS$_2$ QD, we
calculate the polariton-polariton interaction $U_{pl}$ and the
dimensionless parameter $\alpha$ for various detuning $\Delta$ and QD
radius $r_X$ [see Figs.~3(a) and 3(b)]. The polariton-polariton
interaction can be greater than 1~meV for MoS$_2$ for negative
detuning. However, at too large negative detuning, the polaritons
behave like an exciton, and impedes photonic quantum simulation as the
photon addressability of the polaritons is significantly reduced. From
Fig.~3(a) the accessible polariton repulsion can be as large as
several meV. The dimensionless parameter $\alpha$ gives the parameter
regimes for polariton Mott insulator, where the phase boundary is
evaluated again via $\alpha=\alpha_c$ as indicated by the black curve.

\begin{figure}[]
\begin{center}
\includegraphics[height=7.5cm]{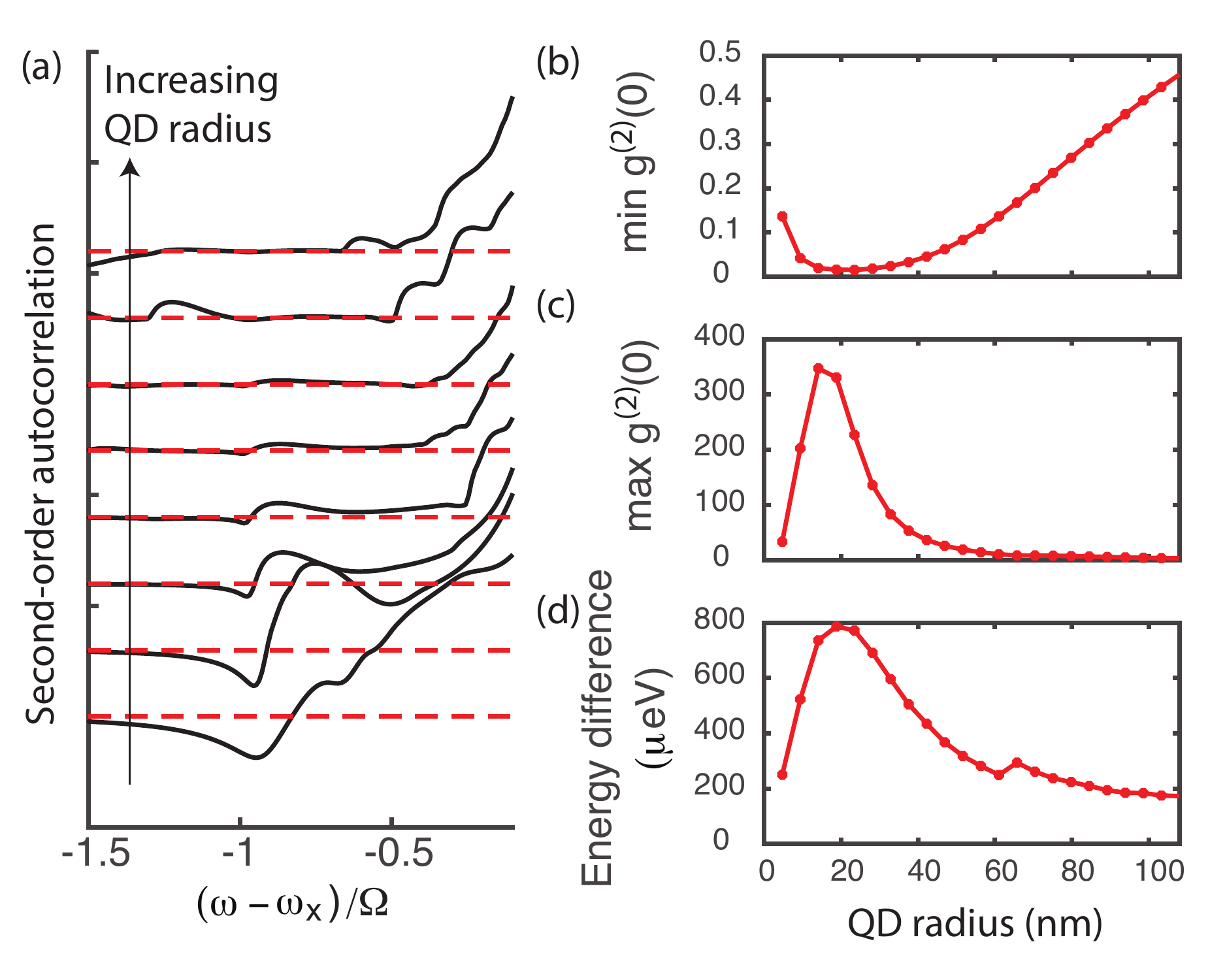}
\caption{ (a) Second-order autocorrelation $g^{(2)}(0)$ vs. optical
  frequency $\ome$ for various QD radius for a MoS$_2$ QD in a single
  cavity as calculated using the Lindblad formalism. The correlation
  function $g^{(2)}(0)$ has a dip (antibunching) at the lower-polariton frequency,
  $\ome_X-\Ome$, and a peak (bunching) on the high-frequency shoulder close to
  the dip. (b) The minimum $g^{(2)}(0)$ at the dip, (c) the maximum
  $g^{(2)}(0)$ on the high-frequency shoulder vs. the QD radius as
  extracted from (a). (d) The frequency difference (as converted to
  energy difference) between the dip and the peak vs. the QD radius.
  The detuning of the quantum emitter and the photon cavity is $\Delta=0$.}
  \label{Fignew}
\end{center}
\end{figure}

To confirm the above findings, we performed full quantum optical
simulation of a tuned single H1-cavity---MoS$_2$-QD hybrid structure,
in presence of the excitonic and photonic losses. In our simulations,
the excitonic loss rate $\gamma_X$ and photonic loss rate $\gamma_c$
are assumed to be same ($\gamma_X=\gamma_c=2\pi$~GHz). We numerically
calculate the evolution of the density matrix by using the standard
Lindblad formalism \cite{AM_NRDC,book-loss}. The calculated
second-order correlation function $g^{(2)}(0)$ shows that the 
photon-antibunching takes place at the lower-polariton frequency $\ome
=\ome_X-\Ome$, corresponding to a dip in the autocorrelation
function $g^{(2)}(0)$ [see Figs.~4(a) and 4(b)]. This is
consistent with both the pictures of photon blockade and polariton
blockade. On the high-frequency shoulder close to the dip, there
always exists a peak of the second-order correlation $g^{(2)}(0)$ [see
Figs.~4(a) and 4(c)]. This photon bunching corresponds to the resonant
excitation of double occupancy of interacting polaritons (i.e.,
adding another polariton to a cavity that already has one
polariton). Therefore, the frequency difference between the peak and
the dip, $\ome_{peak}-\ome_{dip}=\ome_{peak}-\ome_X+\Ome\simeq
U_{pl}/\hbar$, gives a good evaluation of the polariton interaction
strength $U_{pl}$. This frequency difference is indeed maximized for
the QD radius slightly below 20~nm [see Fig.~4(d)]. 
Both the photon antibunching at the dip and the bunching at the peak
become very significant for that optimized QD radius [see Figs.~4(b) and 4(c)].

\section{Quantum many-body simulation in a finite chain of coupled cavities}

The effect of strong interaction between polaritons can be
characterized by the second- and third- order correlation functions
which can be measured experimentally\cite{3rdg}. We have studied such
correlation functions for single cavities in the previous sections. We
now show that these correlations can also extract useful information of the
complex many-body ground state wavefunction of a finite 1D chain of
serially coupled cavities.

\begin{figure}[]
  \begin{center}
    \includegraphics[height=8cm]{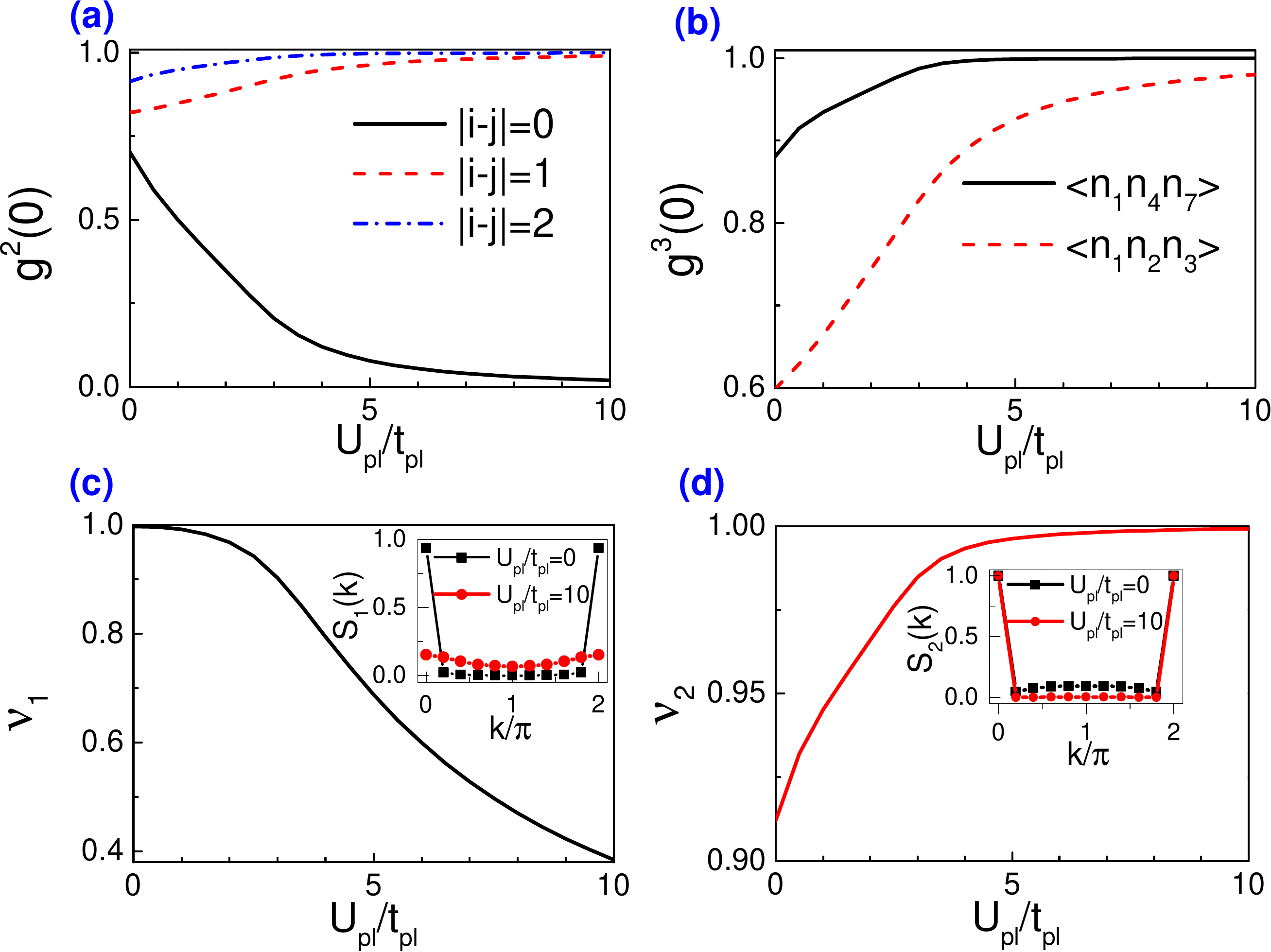}
    \caption{ Second-order and third-order correlation function of the
      ground state of 1D Bose-Hubbard model Eq.~(4). (a) Equal-time
      second-order correlation functions $g^{(2)}(0)$ at the same site
      (black curve) and for nearby sites (red and blue curves)
      vs. $U_{pl}/t_{pl}$. (b) Equal-time third-order correlations for
      nearby sites as functions of $U_{pl}/t_{pl}$. (c) The
      structure factor $S_1(k)$ for photon field and visibility fringe ${\cal V}_1$
      for different $U_{pl}/t_{pl}$. (d) The structure factor $S_2(k)$
      for photon number and visibility fringe ${\cal V}_2$ for various
      $U_{pl}/t_{pl}$.}
    \label{fig3}
  \end{center}
\end{figure}

We calculate the ground state wavefunction of a 1D lattice of
interacting polaritons using exact diagonalization method. We use
periodic boundary condition for $N=10$ sites where each site can have
maximum $3$ polaritons. We note that for a chain of cavities, we did
not consider loss, as a full master equation simulation of the whole
chain is computationally intractable due to extremely large Hilbert
space. While this is a limitation of the present theoretical
treatment, it is the same reason why quantum simulation is highly
sought after. The ground state wavefunction of the system is very
complex. It contains many kinds of long-range multi-particle
entanglement\cite{wen}. A way to characterize such entanglement is to
measure the multi-photon correlations.
We calculate the following correlation functions using the many-body
{\em ground state} obtained from exact diagonalization of the
Bose-Hubbard Hamiltonian (\ref{pol-in}): 
\begin{align}
& g_{ij}^{(2)}(0) = \frac{\langle a^\dagger_i a^\dagger_j a_i
  a_j\rangle}{\ave{a^\dagger_i a_i}\ave{a^\dagger_j a_j}} , \\
& g_{i\ne j\ne l}^{(3)}(0) =
\frac{\ave{n_in_jn_l}}{\ave{n_i}\ave{n_j}\ave{n_l}} ,
\end{align}
where $n_i=a^\dagger_ia_i$. In the regime where $\hbar\Ome\gg t$ the
lower polariton picture is 
well-defined, the $g_{ij}^{(2)}(0)$ correlation function is
proportional to the second-order photon correlation that can be
determined via Hanbury Brown and Twiss measurements.
We calculate $g^{(2)}(0)$ and $g^{(3)}(0)$ for the ground state with
various $U_{pl}/t_{pl}$ (results are shown in Figs. \ref{fig3}(a) and \ref{fig3}(b)). The
second-order correlation function at the same site $g^{(2)}_{ii}(0)$
decreases quickly with increasing $U_{pl}/t_{pl}$, which signifies
photon antibunching due to strong polariton repulsion. On the other
hand $g^{(2)}(0)$ at different sites increases with increasing
$U_{pl}/t_{pl}$, consistent with the understanding that the Mott
insulator state is mostly a product state (plus quantum fluctuations)
with each site occupied by a single polariton. Fig.\ref{fig3}(b) shows
the build-up of $g^{(3)}(0)$ correlations with increasing
$U_{pl}/t_{pl}$ which signifies the localization of polaritons due to 
their mutual repulsion. Those correlation functions reveal the complex
inter-particle entanglement in the strongly interacting polariton
systems which can be sources for nonclassical, highly-entangled
light.

{We also computed the following structure factors 
\begin{align}
& S_1(k) = \frac{1}{N}\sum_{j}\ave{a_i^\dagger a_j}e^{ik(i-j)/N} , \\
& S_2(k) = \frac{1}{N}\sum_{j}\ave{n_i n_j}e^{ik(i-j)/N} ,
\end{align}
as well as the visibility fringes\cite{CCA_1,CCA_2,CCA_3}
\begin{align}
& {\cal V}_1 = \frac{\left . S_1\right|_{max} - \left
    . S_1\right|_{min}}{\left . S_1\right|_{max} + \left
    . S_1\right|_{min}} ,\\
& {\cal V}_2 = \frac{\left . S_2\right|_{max} - \left
    . S_2\right|_{min}}{\left . S_2\right|_{max} + \left
    . S_2\right|_{min}} .
\end{align}
The results are plotted in Figs.~4(c) and 4(d). The visibility fringe
${\cal V}_1$ decreases dramatically with increasing $U_{pl}/t_{pl}$, which
is the signature of the emergence of the Mott insulator state. On the
other hand, the visibility fringe ${\cal V}_2$ increases only slightly with
increasing $U_{pl}/t_{pl}$. The visibility fringe ${\cal V}_1$ has large
contrast for the BEC and Mott insulator states because the BEC is a
coherent state with long range single-particle correlation, while the
Mott insulator is a gapped state with short-range single-particle
correlation.}

\section{Quantum simulation of superlattices of coupled cavity arrays}
\begin{figure}[]
  \begin{center}
    \includegraphics[height=5.cm]{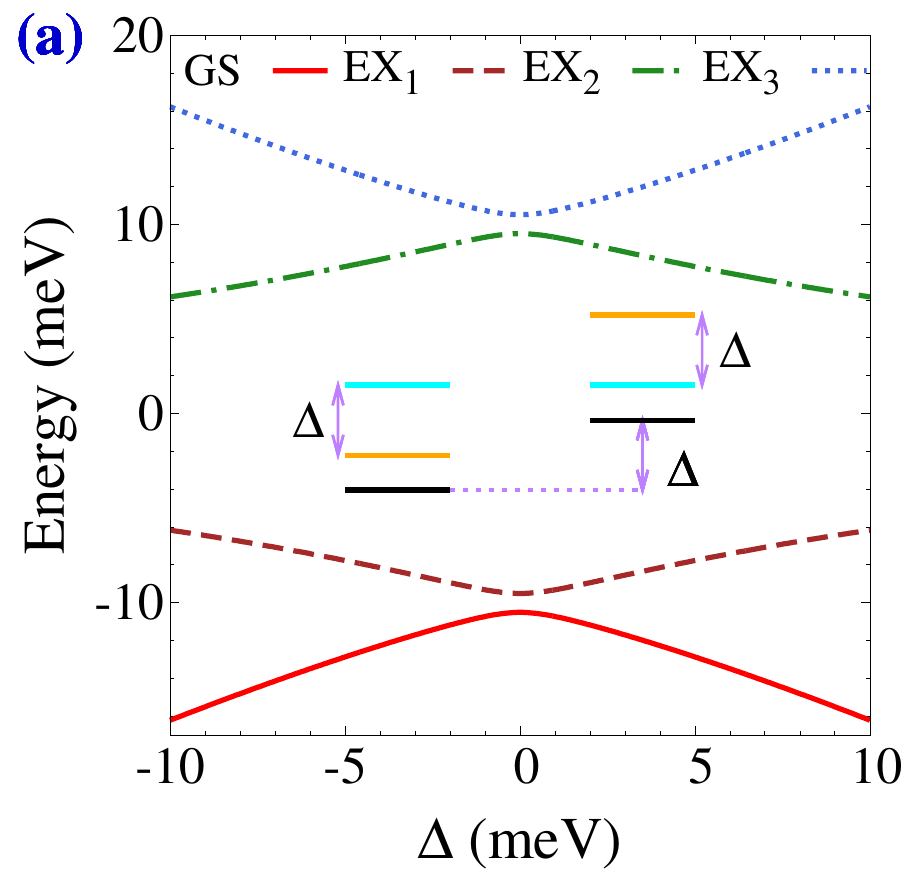}\includegraphics[height=5.cm]{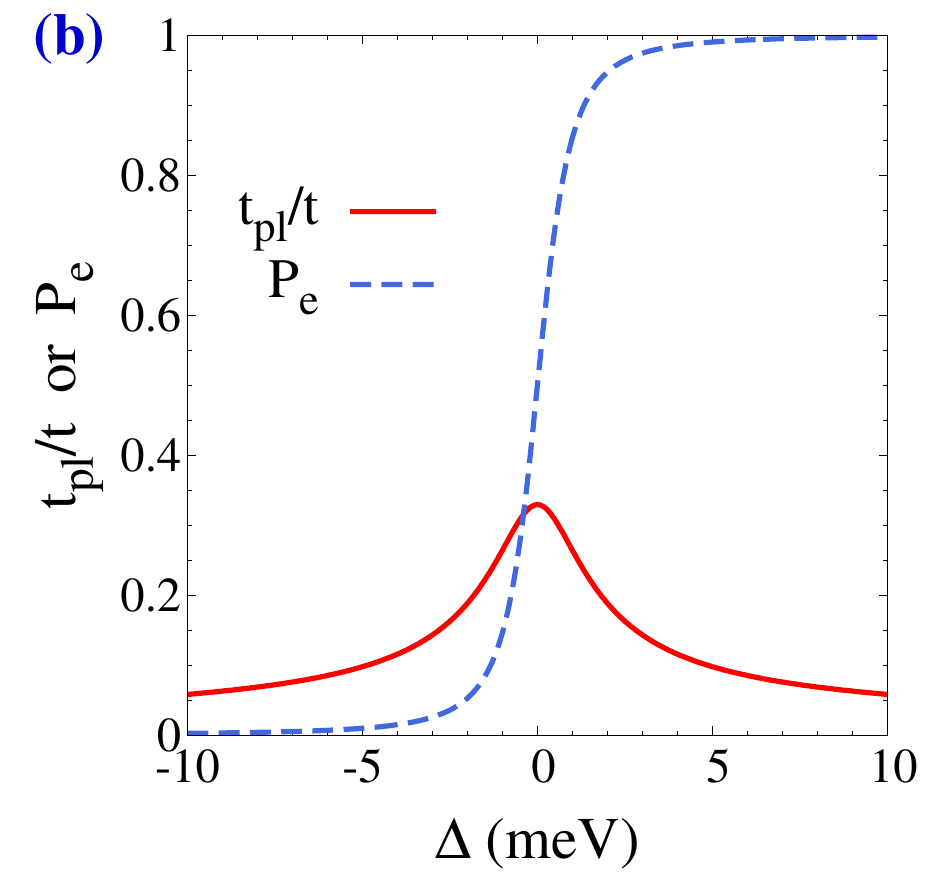}
    \includegraphics[height=5cm]{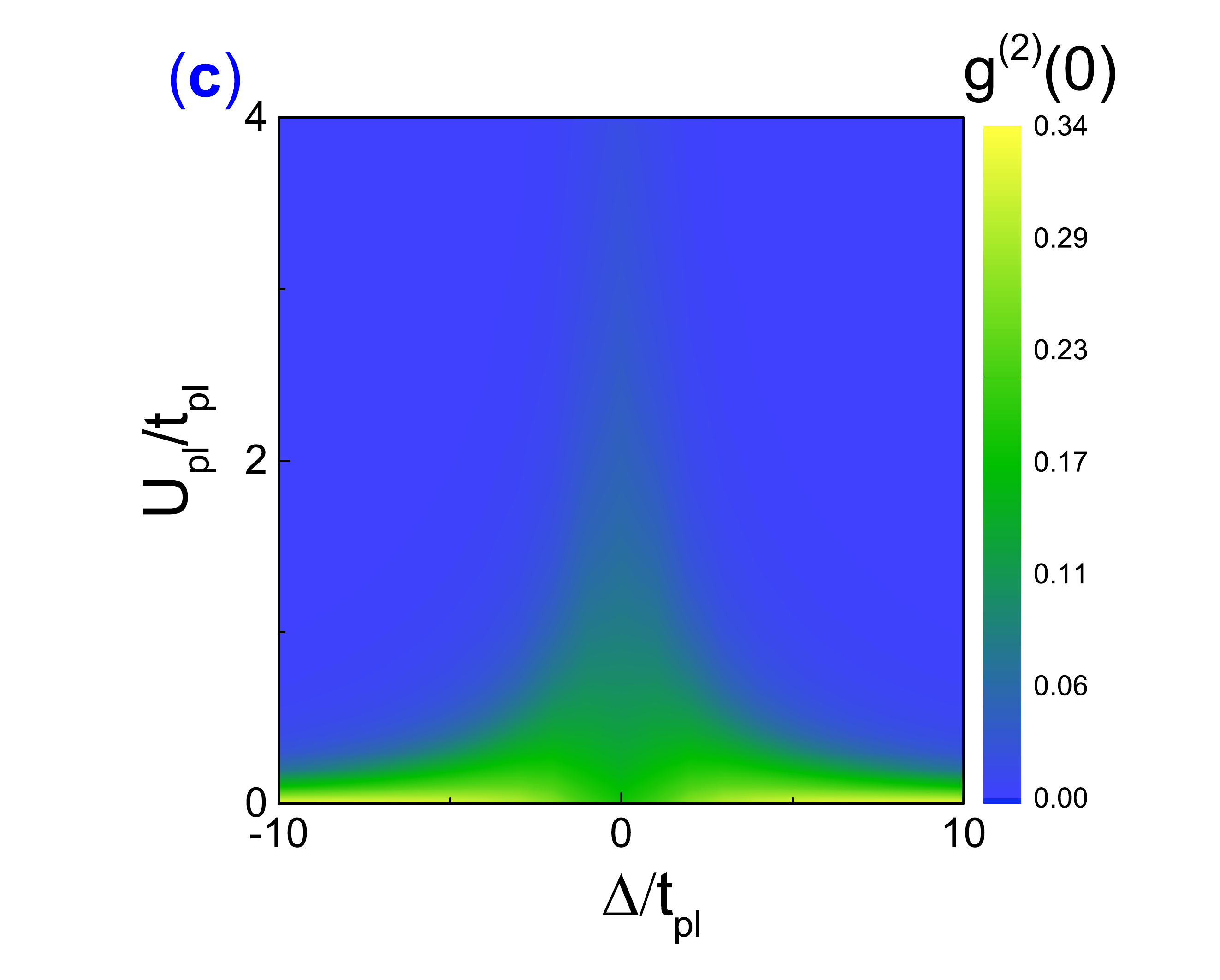}\includegraphics[height=5cm]{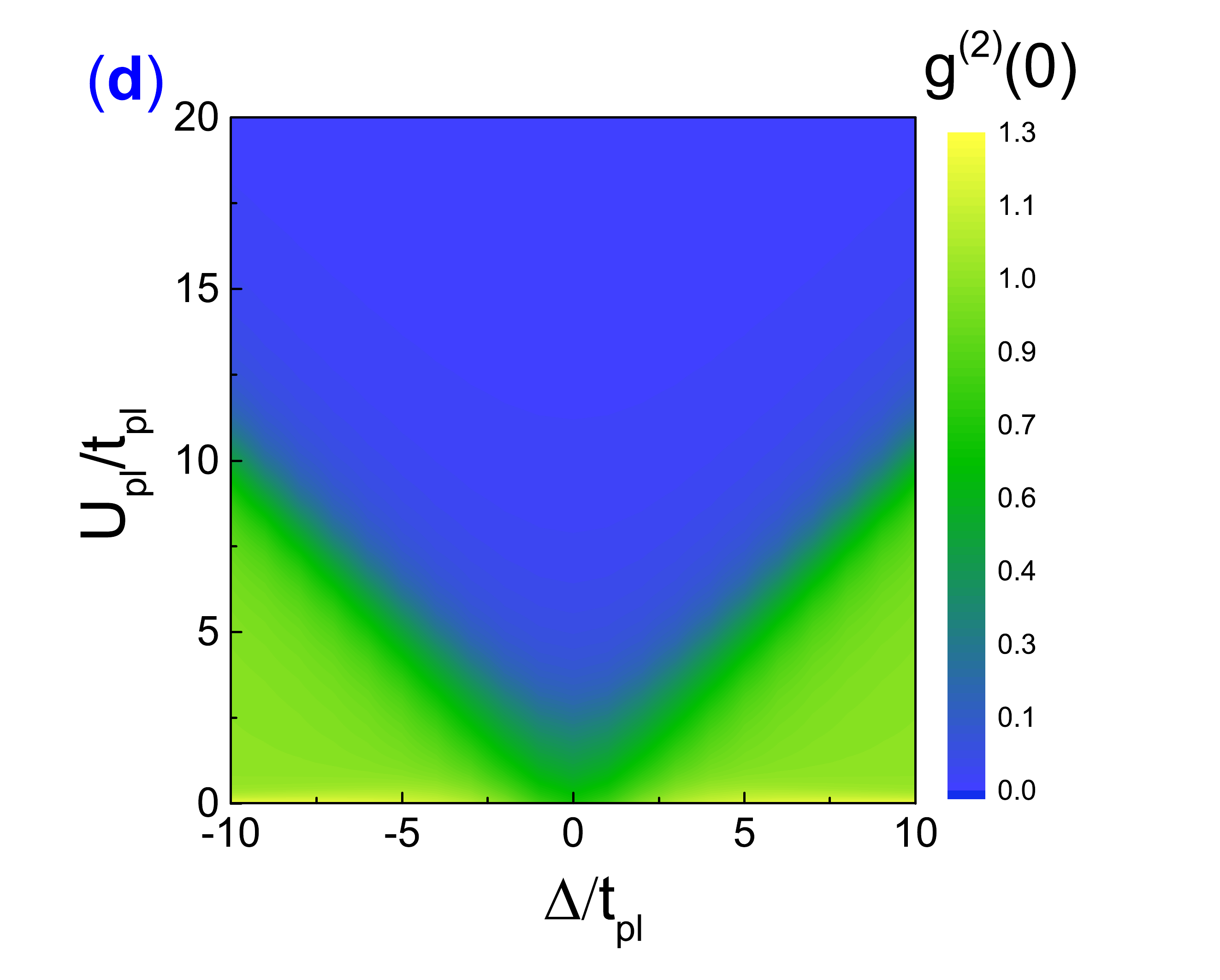}
    \caption{ Quantum simulation in 1D Cavity-QD
      superlattices. (a) Single-particle energy spectrum of coupled double 
      cavity. Inset: Left (right) is the exciton (cyan), cavity (orange),
      lower polariton (black) energy levels for the even (odd) sites. The
      detuning at the even sites is positive, while at the odd sites it
      is negative. The energy difference between lower polariton levels
      in the two different sites is just the detuning $\Delta$. 
      (b) Hopping energy $t_{pl}/t$ and the probability at even site
      $P_e$ as functions of the detuning. Parameters for (a) and
      (b): $t=1$~meV and $\hbar\Ome=10$~meV. (c) The correlation function
      $g^{(2)}(0)$ (averaged over the odd and even sites) as a
      function of the staggered detuning and the interaction strength
      for filling factor $\nu=1/2$ in a finite chain of 10 sites.
      (d) Similar to (c) but with filling factor $\nu=1$.}
    \label{fig4}
  \end{center}
\end{figure}

We now introduce a method for quantum simulation of superlattices of
serially coupled cavity arrays. This can be done via modulating the
detuning, for example, by making the detuning at even (odd) lattice sites as
$\Delta (-\Delta)$. The opposite detunings at two cavities modify
the single particle spectrum and the effect of interaction. The single
particle spectrum of a pair of such detuned cavities is plotted in
Fig. \ref{fig4}(a). The splitting between the ground state and the first
excited state is $\sqrt{\Delta^2+t^2}$. Thus at large detuning
$|\Delta|$ the full Hamiltonian can be truncated into the Hilbert
space of the lowest energy state of a pair of cavity. In this regime
each pair of cavities contribute only one single particle
state (see Fig. \ref{fig4}(b)). Therefore, at half-filling $\nu=1/2$ the
polaritons system can have phase transition into the Mott insulator
state if the interaction between polaritons is strong.

We calculate the many-body ground state of a finite chain of coupled
cavities (with 10 cavities) for filling factor $\nu=1/2$ and
$\nu=1$ by exact diagonalization of the many-body Hamiltonian with a
cut-off of the single site Hilbert space at three bosons. From the
ground state wavefunction, we can obtain the second-order correlation
function $g^{(2)}(0)$ for the even sites and the odd sites. The
averaged $g^{(2)}(0)$ is more relevant to experimental measurements
since it is difficult to distinguish photons from the even site or the
odd site. The significant reduction of the $g^{(2)}(0)$ correlation
function below unity signifies the transition into the Mott insulator
states. For half-filling, the Mott transition is facilitated by the
staggered detuning, which is consistent with the superlattice picture.
For $\nu=1$ filling, the effective filling factor at large detuning $|\Delta|$
is 2. Since the Mott transition at higher filling factor requires
larger interaction strength, the staggered detuning impedes Mott
transition in the large detuning $|\Delta|$ regime for $\nu=1$. Our numerical simulation is
also inconsistent with a few-particle analysis in the Supplemental
Materials.

\section{Realistic considerations on experimental realization}
\label{fluct}

It is known that lack of control over self-assembled QDs
positioning thwarts the scalability of the nonlinear polariton
system. Deterministic fabrication and positioning of the QDs, by
patterning a quantum well like structure can potentially solve this
problem. Unfortunately, such patterning of usual quantum wells
degrades the exciton significantly\cite{QW_degrade}. Monolayer
materials have been proven to be chemically and mechanically stable
and robust\cite{mak,nnano,revtmd}, and can potentially circumvent
these problems of usual optoelectronic materials. 

The difficulties in precisely positioning QDs to the
center of  each cavity are due to incompatibility of the fabrication
method of photonic crystal cavity and that of the QD. Recently new
fabrication methods for MoS$_2$ monolayer QDs was developed where size
and position of QDs can be controlled much more precisely than previous
methods using lithography\cite{lat-het2}. The
main advantages of using MoS$_2$ monolayer QDs is its unique material
compatibility, and robustness against etching (due to its mechanical
and chemical stability)\cite{arka}. Recent works have demonstrated growth of
a large area of monolayer material\cite{large_area}. In practice one can
start with such a large area of monolayer materials, and pattern it to
create an array of quantum dots. The current state-of-the-art
electron-beam technology can fabricate structure reliably with
sub-1~nm accuracy. A significant uncertainty comes from the etching
process, as the lateral etching of the structure is probabilistic, and
creates large non-uniformity. However, delicate fabrication with
electron-beam lithography and etching of photonic crystal cavities
showed an uncertainty of only 10~nm. Monolayer materials provide 
an excellent opportunity, because due to the thinness of the material,
etching them is simple, and does not cause large lateral etching that
degrades the quality of the sample. Hence, fabricating 20~nm radius
quantum dot, and patterning them in an array with periodicity of
$\sim 200$~nm is well within the current fabrication capability. In the
experiment, one can first fabricate the coupled cavity array, and then
transfer the 2D material to the photonic chip. One can perform an
overlay to align the monolayer quantum dots with the cavities. Note
that, current electron-beam technology also provides an overlay
accuracy of 1~nm. As the cavity lateral mode size is significantly
bigger than 1~nm, the fluctuation in exciton-photon coupling due to QD
positioning can be effectively suppressed. Since the etching processes
affect the photonic crystal cavity negligibly, this method also
decouples the correlation between various parameters in our model.

The main dissipation mechanisms in the coupled-cavity-array system
come from the finite exciton and photon lifetimes\cite{prx}. The
state-of-art fabrication technology of photonic crystal cavity has
enabled good control of cavity frequency and very high quality factors
(over 1 million)\cite{noda-review}. With such fabrication technology,
one can have good control of cavity resonance with wavelength
uncertainty below $1$~nm\cite{noda-review}. The finite lifetime due to
exciton non-radiative decay is, however, a major challenge. Note
that recent works have demonstrated good surface passivation to
reduce the non-radiative recombination\cite{science}. These
experimental advancement encourages us to believe that the exciton
nonradiative decay in the QDs can be as long as the
exciton lifetime in the monolayer ($\gtrsim 70$~ps). At sub-1~K
temperature, exciton nonradiative decay is further suppressed,
which is negligible as the resulting exciton linewidth is much
smaller than other energy scales such as $t_{pl}$ and $U_{pl}$
($\sim 1$~meV).

\section{Conclusion and Discussions}
We propose to realize strongly interacting polariton systems based on
MoS$_2$ QD coupled with the H1 photonic crystal cavity. The material
design enables simultaneous realization of strong exciton-photon
coupling and strong exciton-exciton repulsion. This advantage results in
polariton interaction one order of magnitude stronger than 
in the state-of-art single-photon quantum optical systems. The
strongly interacting polariton systems can serve as a platform for
quantum simulation of many-body entanglement and dynamics at the
energy scale of meV and light sources of highly-entangled,
non-classical photons. We discovered that the optimal polariton
interaction is realized near the crossover between photon blockade and
polariton blockade for single-QD in each cavity.

The fluctuation effects may cause difficulties in
realization of quantum phase transition from BEC to Mott insulator.
On the other hand, it was shown that fluctuations in coupled cavity
systems can lead to polaritonic glass phases\cite{CCA_3}. The 
interplay between disorder and interaction effects in localization
of bosonic particles is an interesting physics problem that has been
studied for a long time but unsolved. This regime is also related to
many-body localization which is an area gaining significant attention
recently\cite{mbl}. In the other limit, even a few coupled cavities
\cite{arka} can provide a platform for quantum simulation of strongly
interacting few-particle bosonic systems and serve as multi-photon
entanglement light sources\cite{lurmp}. Finally, we remark that
recent experiments have shown that exciton-exciton interaction can
be tuned via the density of coexisting electrons (or holes) in the
QDs\cite{tune}, offering additional tunability of the system.


\section*{Acknowledgments}
H.X.W and J.H.J acknowledge supports from National Natural Science
Foundation of China (grant no. 11675116) and the Soochow University. J.H.J
also thanks Sajeev John, Gang Chen, and Ming-Qi Weng for
helpful discussions. A.Z
and A.M are supported by the National Science Foundation under grant
NSF-EFRI-1433496; and the Air Force Office of Scientific
Research-Young Investigator Program under grant
FA9550-15-1-0150. A.M. also acknowledges useful discussions with 
Xiaodong Xu. 
W.L.Y acknowledges support by the Natural Science
Foundation of Jiangsu Province of China under Grant No. BK20141190 and
the National Science Foundation of China under Grant No. 11474211. Y.D.X and H.Y.C thank supports from the National
Science Foundation of China for Excellent Young Scientists
(no. 61322504).

{}


\begin{thebibliography}{999}

\bibitem{feynman} K. Kang et. al., Nature (London) {\bf 520}, 656 (2015).


\bibitem{nat} M. Greiner, O.Mandel, T. Esslinger, T. W. Hansch, and
  I. Bloch, Nature (London) {\bf 415}, 39 (2002).

\bibitem{natphys} I. Bloch, J. Dalibard, and S. Nascimbene,
  Nature Phys. {\bf 8}, 267 (2012).

\bibitem{lukin} D. E. Chang, V.Vuleti\'c, and M. D. Lukin, Nat. Photon. {\bf 8},
  685 (2014).

\bibitem{CCA_1} M. J. Hartmann, F. G. S. L. Brand\ ao, and M. B. Plenio,
  Nature Phys. {\bf 2}, 849 (2006).

\bibitem{CCA_2} D. G. Angelakis, M. F. Santos, and S. Bose,
  Phys. Rev. A. {\bf 76}, 031805(R) (2007).
 

\bibitem{CCA_3} D. Rossini and R. Fazio,
  Phys. Rev. Lett. {\bf 99}, 186401 (2007).

\bibitem{CCA_4} C. Noh and D. G. Angelakis, Rep. Prog. Phys. {\bf 80}, 016401 (2016).

\bibitem{ciutirmp} I. Carusotto and C. Ciuti,
  Rev. Mod. Phys. {\bf 85}, 299 (2013).

\bibitem{jelena} S. Buckley, K. Rivoire, and J. Vu\v{c}kovi\'c,
  Rep. Prog. Phys. {\bf 75}, 126503 (2012).

\bibitem{polariton} T. Byrnes, N. Y. Kim, and Y. Yamamoto,
  Nature Phys. {\bf 10}, 803 (2014).  

\bibitem{3rdg} A. Rundquist {\sl et al.}, Phys. Rev. A {\bf 90}, 023846 (2014). 

\bibitem{wen} X. G. Wen, {\it Quantum Field Theory of Many-Body
  Systems} (Oxford University Press, 2004).

\bibitem{prx} J.-H. Jiang and S. John,
  Phys. Rev. X {\bf 4}, 031025 (2014).

\bibitem{cdte} J. Kasprzak {\sl et al.}, Nature (London) 
  {\bf 443}, 409 (2006).

\bibitem{snoke} R. Balili, V. Hartwell, D. Snoke, L. Pfeiffer, and K. West,
  Science {\bf 316},
  1007 (2007).

\bibitem{sr} J. H. Jiang and S. John, Sci. Rep. {\bf 4}, 7432 (2014).

\bibitem{diode} H. S. Nguyen {\sl et al.}, Phys. Rev. Lett. {\bf 110}, 236601
  (2013).

\bibitem{transistor} D. Ballarini {\sl et al.}, Nat. Commun.
  {\bf 4}, 1778 (2013).

\bibitem{verger} A. Verger, C. Ciuti, and I. Carusotto,
  Phys. Rev. B {\bf 73}, 193306 (2006).

\bibitem{pblock0} K. M. Birnbaum, A. Boca, R. Miller, A. D. Boozer, T. E. Northup,
  and H. J. Kimble, Nature (London) {\bf 436}, 87 (2005).

\bibitem{pblock1} A. Faraon, I. Fushman, D. Englund, N. Stoltz, P. Petroff, and
  J. Vu\v{c}kovi\'c, Nature Phys. {\bf 4}, 859 (2008).

\bibitem{pblock2} A. Reinhard, T. Volz, M. Winger, A. Badolato, K. J. Hennessy,
  E. L. Hu, and A. Imamo\ifmmode~\bar{g}\else\={g}\fi{}lu, Nat. Photon. {\bf 6},
  93 (2012).

\bibitem{ele2} K. M\"uller {\sl et al.}, Phys. Rev. Lett.
  {\bf 114}, 233601 (2015).


\bibitem{ourcavity} J. Hagemeier {\sl et al.}, Opt. Exp.
  {\bf 20}, 24714 (2012).

\bibitem{SM} See Supplementary Materials, http://


\bibitem{nc-mse2}
S. Dufferwiel
  {\sl et al.}, Nat. Commun.
  {\bf 6}, 8579 (2015).

\bibitem{diana}
D. Y. Qiu, F. H. da Jornada, and S. G. Louie,
Phys. Rev. Lett. {\bf 111},
216805 (2013).


\bibitem{tassone}
F. Tassone and
  Y. Yamamoto,
  Phys. Rev. B {\bf 59},
  10830 (1999).




\bibitem{PhysRevB.58.R14741} T. D. K\"uhner and
  H. Monien,
  Phys. Rev. B {\bf 58},
  R14741(R) (1998).

  \bibitem{AM_NRDC}
  A. Majumdar {\sl et al.},
  Phys. Rev. B {\bf 84},
  085309 (2011).


\bibitem{book-loss} C. Gardiner and P. Zoller, {\it Quantum Noise},
  3rd ed. (Springer, Berlin, 2004).




















\bibitem{QW_degrade} T. C. Weisbuch, R. Dingle, A.C. Gossard, and W. Wiegmann,
    Solid State Comm. {\bf 38}, 709-712 (1981).



\bibitem{mak}
  K. F. Mak,
  C. Lee,
  J. Hone,
  J. Shan, and
  T. F. Heinz,
  Phys. Rev. Lett. {\bf 105},
  136805 (2010).


\bibitem{nnano}
Q. H. Wang,
  K. Kalantar-Zadeh,
  A. Kis,
  J. N. Coleman,
  and M. S.
  Strano, Nat. Nanotech.
  {\bf 7}, 699 (2012).

\bibitem{revtmd}
X. Xu,
  W. Yao,
  D. Xiao, and
  T. F. Heinz,
  Nature Phys. {\bf 10},
  343 (2014).

\bibitem{lat-het2} G. Wei {\sl et al.},
  arXiv: 1510.09135.


\bibitem{arka}
  S. Wu {\sl et al.},
  Nature (London) {\bf 520},
  69 (2015).

\bibitem{large_area} D. Dumcenco {\sl et al.}, ACS Nano, {\bf 9},
  4611 (2015).
   

\bibitem{noda-review}
S. Noda,
  M. Fujita, and
  T. Asano,
  Nat. Photon. {\bf 1},
  449 (2007).

\bibitem{science} M. Amani {\sl et al.}, Science 
  {\bf 350}, 1065 (2015).


\bibitem{mbl} V. Oganesyan and D. A. Huse, Phys. Rev. B {\bf 75},
  155111 (2007); A. Pal and D. A. Huse, Phys. Rev. B {\bf 82}, 174411
  (2010); Ehud Altman and Ronen Vosk, Ann. Rev. Cond. Matt. Phys. {\bf
    6}, 383 (2015).


\bibitem{lurmp} J.-W. Pan, Z.-B. Chen, C.-Y. Lu, H. Weinfurter,
  A. Zeilinger, and M. Zukowski, Rev. Mod. Phys. {\bf 84},
  777 (2012).

\bibitem{tune} M. Sidler {\sl et al.}, arXiv:1603.09215






\end{thebibliography}
\end{document}